\documentclass{article}
\usepackage {graphicx}

\title{Excitability in a Model with a Saddle-Node Homoclinic
      Bifurcation}\label{title}
\author{Rui Dil\~ao
        and
        Andr\'as Volford}
\date{}

\begin{document}

\maketitle

\centerline{\scshape   Rui Dil\~ao}
 \medskip

  {\footnotesize \centerline{Nonlinear Dynamics Group, Instituto Superior T\'ecnico}
  \centerline{Department of Physics, Av. Rovisco Pais, 1049-001 Lisbon, Portugal} }
 \medskip

\centerline{\scshape  Andr\'as Volford}
 \medskip

 {\footnotesize \centerline{Center for Complex and Nonlinear Systems,
  Technical University of Budapest}
   \centerline{H-1521 Budapest, Hungary} }

 \medskip

\begin{abstract}
 In order to describe excitable reaction-diffusion systems, we derive a two-dimensional model with a Hopf and a semilocal saddle-node homoclinic bifurcation. This model gives the theoretical framework for the analysis of the saddle-node homoclinic bifurcation as
observed in chemical experiments, and for the concepts of excitability and excitability threshold.
We show that if diffusion drives an extended system across the excitability threshold
then, depending on the initial conditions, wave trains, propagating solitary pulses and propagating pulse packets can exist in the same extended system.
The extended model shows  chemical turbulence for equal diffusion coefficients and presents
all the known types of topologically distinct activity waves observed in chemical experiments. In particular, the approach presented here
enables to design experiments in order to decide between excitable systems with sharp and finite width thresholds.
\end{abstract}

\section{Introduction}

A system is called extended if it consists of many individual similar subsystems distributed in space, and all the subsystems have the same dynamics. Examples of extended systems are obtained by assembling similar subsystems through diffusive coupling. If all the local subsystems are in the same state, that is, synchronized, the extended system behaves globally as the local subsystem. However, if one of the subsystems is desynchronized, we can have the emergence of collective behavior, and the dynamic behavior of the extended system is dependent on both the initial spatial distribution of the desynchronized subsystems and the parameters of the local system.

Here we are interested in the emergence of collective behavior of an extended system when  the initial conditions of a  local subsystem are changed.
When the local dynamics has a stable steady state in the phase space and is close to a homoclinic orbit, we show that, in the extended system, wave trains, propagating solitary pulses and propagating pulse packets can develop.
These types of wave patterns depend on the distribution of the initial states
of the extended system.

An extended system is excitable if after being slightly locally perturbed from a stable steady state it returns to the quiescent state, but shows traveling activity waves if perturbations are large enough. Systems responding in this way to perturbations are said to have a threshold.  Systems with excitable behavior appear in extended chemical reactions \cite{1}, in the propagation of
action potentials along nerve axons \cite{2,3}, and in population dynamics models \cite{4,h}.

In the context of extended chemical reactions,
there are two main types of models showing excitability. One of these models, the nullcline model, has been proposed by Tyson and Fife \cite{1} and Tyson and Keener \cite{5,6} to explain excitability effects in chemical experiments. In this model, excitability  is related to the particular behavior of the nullclines in the two-dimensional phase space of a  differential equation: one nullcline is N-shaped, the other is monotonic and they intersect at only one point. Excitability develops as a continuous process when initial conditions are changed across the N-shaped nullcline,
leading to an abrupt continuous change of the orbit length in phase space.
In this case, the system does not have a precise threshold, but a finite narrow threshold region. These models of excitable systems  have only one stable fixed point
and are usually associated with two-variable models, with a slow and a fast dynamics \cite{7}. In the literature of excitable systems they are called Type I excitability models.

Another model for excitable behavior has been proposed by Noszticzius, Witt\-mann and Stirling \cite{8}, McCormic,  Noszticzius and Swinney \cite{9} in the context of the observed behavior in Belousov-Zhabotinskii type chemical systems, and by Ermentrout and Rinzel \cite{10} in the context of propagation of action potentials along nerve axons. This model gives a qualitative explanation for the appearance of an excitable steady state in chemical experiments and in the propagation of action potentials along nerve axons through a bifurcation in the local kinetics, from a limit cycle to a saddle-node pair of fixed points. Some authors called it saddle-node infinite period or SNIPER model \cite{9}, or Type II excitability model. This qualitative model has three fixed points in phase space, one saddle point, one stable node and one unstable focus. In this case, the phase portrait has a sharp threshold of excitability associated with
the stable manifold of the saddle point. The threshold of excitability has two branches, one connecting a saddle point with the unstable focus, and the other starting at the saddle point and going to infinity.
Several reaction-diffusion models with these features appeared in the literature \cite{11,12}. These excitability models can produce propagating pulses in extended media as observed in experiments \cite{13}.

The SNIPER or saddle-node homoclinic (SNH) bifurcation has
been predicted by Leontovich \cite{16} and Andronov {\it et al.} \cite{17}, as a topological possibility for the break of
hyperbolicity of limit cycles in two-dimensional continuous dynamical systems, and is one of the simplest codimension-1 semilocal bifurcations \cite{18}.
As the SNH  bifurcation is nonlocal,  local analysis describes the local
saddle-node bifurcation, missing the appearance of the limit cycle \cite{19}. Therefore, it is difficult
to build a sufficiently general explicit analytical model for this global bifurcation.

For several chemical systems in stirred tank reactor experiments, Maselko and Epstein \cite{20} have constructed qualitative experimental phase portraits compatible with the experimental behavior and a saddle loop bifurcation scenario. In this approach, the qualitative
phase portraits have one stable and two unstable fixed points, but are obtained
through a saddle loop codimension-2 (Bogdanov-Takens) bifurcation of a fixed point \cite{21}. To have bounded motion in phase space, we must
have an additional stable node external to the saddle loop. However, the
phase portraits of these systems are topologically equivalent to those obtained through the saddle-node homoclinic (SNH) bifurcation. Hence, an excitability model with threshold can be obtained by two
different  bifurcation scenarios, but at the excitable regime,
they have the same phase space structure. In the following, we adopt the simplest  bifurcation scenario through the codimension-1 SNH bifurcation.

Our aim here is to derive a sufficiently general excitability model compatible with the above mentioned
experimental results. We begin by constructing an explicit model that exhibits a Hopf and a saddle-node homoclinic bifurcation, both driven by independent parameters. Then,  we add diffusion to
the two phase space coordinates, obtaining a general model of a reaction-diffusion
excitable media. With a bench-marked numerical method, we derive properties of the solutions of the one- and the two-dimensional reaction-diffusion systems.
The Hopf bifurcation will be essentially associated with the existence of stable oscillations of the local kinetics, as observed in continuous stirred tank reactor  experiments \cite{23}.
The saddle-node homoclinic bifurcation will be associated with the break up of the periodic behavior of the local kinetics and the appearance of a new steady state with the excitability property \cite{8}.

We show that, at and  beyond the SNH bifurcation, an extended chemical system coupled by diffusion develops traveling wave trains, propagating pulses, pulse packets and fronts, and aperiodic wave trains or turbulent states.
Periodic wave trains, pulses and pulse packets are tuned by different initial conditions. Turbulent states and fronts are obtained by changing the parameters of the local model. All these
situations occur for equal diffusion coefficients in the two phase space variables.

In the next section, we derive the main properties of locally excitable two-dimensional models associated with the SNH bifurcation, we derive its geometric features in phase space, and we characterize its bifurcation set. In section 3, the diffusion coupling is introduced, and the properties of the solutions of the associated one- and two-dimensional reaction-diffusion system are analyzed by numerical methods. As we are dealing with a parabolic partial differential equation, we emphasize our analysis on the pattern formation features associated with the initial conditions of the extended system. Finally, in section 4 we discuss the main conclusions of the paper.

\section{The saddle-node homoclinic (SNH) bifurcation}

In order to construct an analytical model exhibiting a Hopf and a SNH bifurcation, the starting point will be the versal unfolding of the Hopf bifurcation in polar coordinates, \cite{21}:
\begin{eqnarray} \label{E:1}
\frac{d r}{d t} &=& r(\nu  + a r^2 )\nonumber\\
\frac{d \theta}{d t} &=& \beta  + b r^2
\end{eqnarray}
where $a$, $b$ , $\nu$ and $\beta $ are real parameters. In the literature of chemical kinetics, equation (\ref{E:1}) is the Stuart-Landau equation with real parameters \cite{24}.

If, $a<0$, $\beta \not= 0$,  and $\nu >0$,  equation (\ref{E:1}) has one limit cycle (in Cartesian coordinates) in phase space and an unstable focus at the origin. The radius of
the limit cycle is $r_0=\sqrt{-\nu/a}$. For, $a<0$, $\beta\not =0$, and $\nu=0$, the
system (2.1) has a supercritical codimension-1 Hopf bifurcation.
For, $a<0$, and $\nu<0$, the origin in phase space is a stable focus. If $\beta =0$ and $\nu =0$, we
are in the conditions of a (Bogdanov-Takens) codimension-2 bifurcation \cite{21}.
In this case, (\ref{E:1}) is no longer a versal unfolding for this bifurcation.

One of the features of the normal form (\ref{E:1}) is that the phase equation is phase independent. From the analysis of the vector field  (\ref{E:1}), for $a<0$, and $\nu>0$, in Cartesian coordinates, the rotation frequency of orbits in phase space near the origin approaches $\beta $. On the limit cycle, the rotation frequency is $\omega= \beta+br_0^2= \beta-b\nu/a$. If, $a<0$, $\beta >0$, and $b>\beta a/\nu$, rotations near the origin and in the limit cycle are counter-clockwise. If, $a<0$, $\beta >0$, but $b<\beta a/\nu$, the rotation near the limit cycle becomes clockwise. For, $a<0$, $\beta>0$, and $b=\beta a/\nu$, the rotation frequency on the limit cycle becomes zero, all the points on the limit cycle are fixed points, and the phase space trajectories on the limit cycle do not rotate. Under this condition ($b=\beta a/\nu$),  we have a line of fixed points. Due to this property of the normal form  (\ref{E:1}), if we make the parameter $b$  phase dependent, we can incorporate a saddle-node bifurcation into the limit cycle, obtaining a non-local, SNH, bifurcation for the full dynamics. So, we change the system of equations (\ref{E:1}) to the new system,
\begin{eqnarray}\label{E:2}
\frac{d r}{d t} &=& r(\nu  + a r^2 )  \nonumber\\
\frac{d \theta}{d t} &=& \beta  + (b  + \alpha f(\theta ))r^2
\end{eqnarray}
where $\alpha $ is a new parameter, and $f(\theta )$ is a  non-negative function that introduces a phase dependence on the phase equation. To maintain
the continuity of trajectories in the two-dimensional phase space, we consider
that $f(\theta )$ is continuous and non-constant, and $f(0)=f(2\pi )$. To simplify the stability analysis, we further consider that $f(\theta )$ has continuous derivatives.

To derive the qualitative structure in phase space of the orbits of the differential equation (\ref{E:2}), we consider the equations for the fixed points,
\begin{eqnarray}\label{E:3}
   r(\nu  + a r^2 ) &=& 0  \nonumber\\
   \beta  + (b  + \alpha f(\theta ))r^2  &=& 0 .
\end{eqnarray}
If, $a<0$, $\nu>0$,  and  $\beta  + (b  + \alpha f(\theta ))r^2>0$ or  $\beta  + (b  + \alpha f(\theta ))r^2<0$, for all $\theta \in [0,2\pi ]$ and $r \in [0,r_0 ]$, then system  (\ref{E:3})  in Cartesian coordinates has one stable limit cycle and the origin is an unstable focus. As in the normal form case, the supercritical Hopf bifurcation occurs for $\nu=0$, provided $\beta \not=0$.

If, $a<0$, $\nu>0$, and  $\beta  + (b  + \alpha f(\theta ))r_0^2=0$, for some  $\theta_1,\theta_2,\ldots \in [0,2\pi ]$, with $f'(\theta_i) \not=0$, then the  system (2.2) has an even number of fixed points with coordinates $(r_0,\theta_1)$,
$(r_0,\theta_2)$, ... .  The Jacobian matrix of equation (\ref{E:2}) at $(r_0,\theta_i)$ is,
\begin{equation}\label{E:4}
J=\pmatrix{
-2\nu & 0\\  2 r_0(b+\alpha f(\theta_i))& -\frac{\alpha  \nu}{ a}f'(\theta_i) } 
\end{equation}
The eigenvalues of the Jacobian matrix (\ref{E:4}) are $\lambda_r=-2\nu$ and
$\lambda_{\theta}=-\alpha  \nu f'(\theta_i)/a$, corresponding to the eigenvectors,
\begin{eqnarray}\label{E:5}
e_{\lambda_r}&=&(\alpha  \nu f'(\theta_i)/a-2\nu)e_{r}+2 r_0(b+\alpha f(\theta_i)) e_{\theta} \nonumber\\
e_{\lambda_{\theta}}&=&e_{\theta} .
\end{eqnarray}
For, $a<0$, and
$\nu >0$,
the angular direction $e_{\lambda_{\theta}}$, taken at these fixed points, is tangent to the circle of radius $r_0$, which, by  (\ref{E:3}), is an invariant set in phase space. Therefore,
the circle $r=r_0$  is the union of the stable and unstable manifolds connecting
the  fixed points with coordinates $(r_0,\theta_i)$. According to the sign of the derivatives $f'(\theta_i)$, the  fixed points $(r_0,\theta_i)$ are alternately saddle points and stable nodes, along the circle $r=r_0$. If $f'(\theta_i) >0$, then $(r_0,\theta_i)$ is a saddle. By  (\ref{E:5}), if
$(\alpha  \nu f'(\theta_i)/a-2\nu) \not=0$ the stable manifolds associated
with the saddle points are tangent to the directions $e_{\lambda_r}$, being transversal to the circle $r=r_0$, and make heteroclinic connections with the origin of coordinates.

Solving the second equation in (\ref{E:3}) for $f(\theta )$, we obtain,
$f(\theta) = -(\beta +b r_0^2)/(\alpha r_0^2)$, and there exists always a
choice of $\alpha $ such that a new fixed point appears at $r=r_0$. As $f(\theta )$ is continuous and non-constant in the interval $(0, 2\pi )$, the new fixed point
appears for,
\begin{equation}\label{E:6}
\alpha_S = -\frac{\beta +b r_0^2}{r_0^2 \max_{\theta } f(\theta)  }
\end{equation}
or,
\begin{equation}\label{E:7}
\alpha_S = -\frac{\beta +b r_0^2}{r_0^2 \min_{\theta } f(\theta)  }
\end{equation}
where the maximum and minimum are taken around the local extremes of $f$. For example, if $f(\theta )$ has only one local maximum in $[0,2\pi ]$, the new fixed point appears for $\alpha =\alpha_S$ given by equation (\ref{E:6}). If $f(\theta )$ has only one minimum,  the new fixed point appears for $\alpha =\alpha_S$ given by equation (\ref{E:7}).

The appearance of these  fixed points on the limit
cycle, when we vary the parameter $\alpha $ across the value $\alpha_S$, is the SNH bifurcation. At the SNH bifurcation, we have locally in phase space one or several saddle-node bifurcations depending of the number of local maxima and minima of
$f(\theta )$.

The SNH bifurcation is tuned by the parameter $\alpha $. For example,
choosing the  function  $f(\theta )=\sin^2(\theta/ 2)$, and
solving the equations for the fixed points, (\ref{E:3}), we obtain,
\[
\alpha  \sin^2(\theta/ 2)=  - \frac{\beta}{{r_0^2 }} - b =  \frac{\beta  a}{ \nu} - b
\]
and  the SNH bifurcation occurs for,
\begin{equation}\label{E:8}
\alpha_S =    \frac{\beta  a}{ \nu} - b
\end{equation}
where one non-hyperbolic fixed point appears and has polar coordinates $(r_0,\pi)$.
 For, $\alpha <\alpha_S$, there are  two new fixed points and their
stability properties are calculated from the eigenvalues of the Jacobian matrix (\ref{E:4}).

\begin{figure}[htbp]
\centerline{\includegraphics[width=3.62in,height=3.56in]{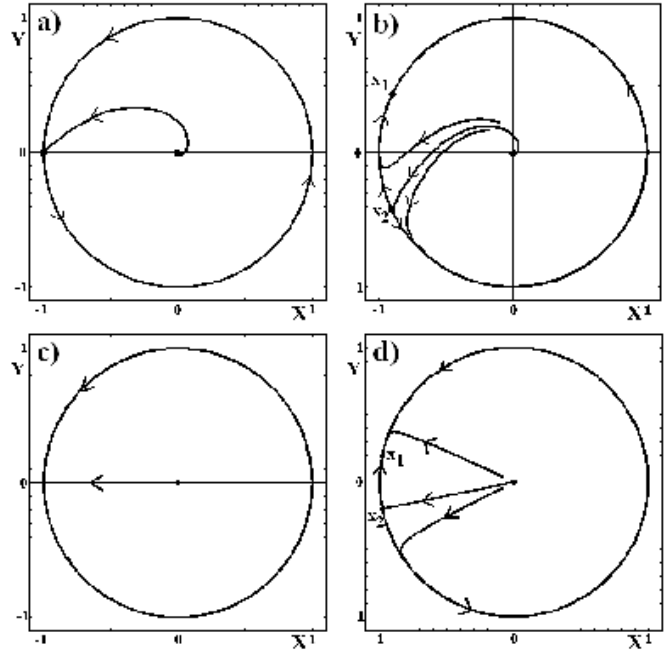}}
\caption{Phase space structure of the differential equation (\ref{E:9}) at the SNH bifurcation (a), and after the bifurcation (b). The parameter values are: $a= -1$, $b=1.0$, $\nu=1$, and in a) $\alpha= -2.0$ and $\beta=1$; b) $\alpha= -2.1$ and $\beta=1$.
According to (\ref{E:8}), the SNH bifurcation occurs for $\alpha_S=-2.0$, corresponding to case a).
Parameter values for cases c) and d) are: c) $\alpha= -1.0$ and $\beta=0$;
d) $\alpha= -1.05$ and $\beta=0$.
The fixed point $x_1$ is a stable node and $x_2$ is a saddle point. The threshold of excitability coincides with the stable manifold of the saddle point, which makes a heteroclinic connection with the origin of coordinates.}\label{f1}
\end{figure}

Changing coordinates from polar to Cartesian, $X=r \cos (\theta )$ and $Y=r \sin (\theta )$, and with $f(\theta )=\sin^2(\theta/ 2)$, system (\ref{E:2}) can be written as,
\begin{eqnarray}\label{E:9}
 \frac{d X}{d t} &=& \nu X - \beta Y \nonumber\\
&+& (X^2  + Y^2 )\left(a X - (b  + \frac{\alpha }{ 2}(1 - \frac{X}{\sqrt {X^2  + Y^2 }}) )Y\right)  \nonumber\\
\frac{d Y}{d t} &=& \beta X + \nu Y \nonumber\\
&+& (X^2  + Y^2 )\left(a Y + (b  + \frac{\alpha}{ 2}(1 - \frac{X}{\sqrt {X^2  + Y^2 }}) )X\right) .
\end{eqnarray}
At the SNH bifurcation, system (\ref{E:9}) has one fixed point away from the origin
of coordinates, $(X=-r_0,Y=0)$.
This fixed point is non-hyperbolic and the circle of radius $r_0$ is a one-dimensional center manifold. The stable manifold of this fixed point makes a heteroclinic connection with the unstable fixed point at the origin of coordinates. The fixed point $(X=-r_0,Y=0)$ is semistable, Fig.~\ref{f1}a).
For,  $a<0$, $\beta\not= 0$, $\nu >0$, and $\alpha < \alpha_S$,
system (\ref{E:9}) has one unstable focus at the origin, and a saddle-node pair of fixed points on the circle $r=r_0$, Fig.~\ref{f1}b). The stable manifold of the saddle point makes a heteroclinic connection with the unstable focus at the origin of coordinates. For,
$a<0$, $\nu >0$, and $\alpha >\alpha_S$, system (\ref{E:9}) has a stable limit cycle in phase space.

At the SNH bifurcation and beyond it, the shape of the heteroclinic connection between the saddle point
and the origin of coordinates can be controlled, independently of the radius
of the invariant circle in phase space. By (\ref{E:5}) and (\ref{E:3}), the tangent space to the stable manifold
of the semistable fixed point at $\alpha =\alpha_S$ is generated by the vector,
\begin{equation}\label{E:10}
e_{\lambda_r}=-2\nu e_{r}-\frac{2\beta}{ r_0}  e_{\theta}
\end{equation}
and the angle between the tangent space to the stable manifold and the tangent space to the center manifold is,
\begin{equation}\label{E:11}
a_{sc}=\arctan{\frac{r_0\nu}{\beta}} .
\end{equation}
For
$\beta =0$, $a_{sc}=\pi /2$,  and the negative part of the $X$-axis is invariant for the flow (\ref{E:9}), Fig.~\ref{f1}c), and the heteroclinic connection coincides with the interval $[-r_0,0]$ of the $X$-axis.  In Fig.~\ref{f1}d), the topological structure of phase space for $\alpha <\alpha_S$ and $\beta =0$ is shown. Increasing $\beta $,
the heteroclinic connection winds around the origin of coordinates in phase space.

\begin{figure}[htbp]
\centerline{\includegraphics[width=2.91in,height=2.0in]{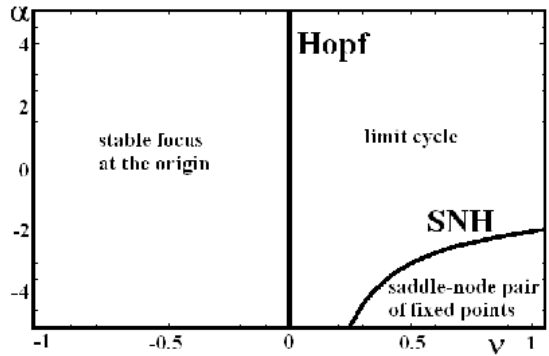}}
\caption{Bifurcation diagram of the Hopf and SNH bifurcations for
equation (\ref{E:9}), for parameter values $a=-1$, $b=1$ and $\beta =1$. The SNH
bifurcation curve is given by (\ref{E:8}).}\label{f2}
\end{figure}

In Fig.~\ref{f2}, we show  the bifurcation diagram for equation (\ref{E:9}), for the co\-di\-men\-sion-1 Hopf and SNH bifurcations,
with $\beta >0$.

An important property of the system of equations (\ref{E:9}) is that, away from the Hopf ($\nu=0$), SNH ($\alpha=\alpha_S$)
and Bogdanov-Takens ($\nu=0$, $\beta =0$)
singularities, all the fixed points are hyperbolic and there are  no saddle-to-saddle connections.
For $\alpha >\alpha_S$, the limit cycle is also a hyperbolic closed orbit. These characteristics of the flow imply that, away from bifurcations, the generic models (\ref{E:2}) and (\ref{E:9}) are structurally
stable in the $C^1$ norm. From the applied point of view, this is equivalent to say that this  model is robust. If $\beta \not=0$, the SNH bifurcation is one of the three possible  semilocal codimension-1 bifurcations of two-dimensional vector fields, and is unfolded by the parameterization
of the vector field (\ref{E:9}) in the sense that when we vary the parameter $\alpha $ across
$\alpha_S$, on both sides of the bifurcation, we have structurally stable (Morse-Smale)
dynamical systems \cite{18}, Fig.~\ref{f2}. If $\beta =0$ and $\nu=0$, we are in a scenario
that corresponds to a Bogdanov-Takens codimension-2 singularity of the origin, but equation (\ref{E:9}) do
not unfold it \cite{21}.

The differential equation (\ref{E:9}) is an analytical model that makes precise the concepts
of excitability  and threshold of excitability as defined by Winfree \cite{25}:
``A reaction is excitable if it has a unique steady state that the system will approach from all the initial conditions, but there exists a locus of initial conditions near which either of two quite different  paths may be taken  toward the unique steady state. If one of these paths is a lot longer than the other, then the system is excitable."
The threshold of excitability is defined by the stable manifold of the saddle point.

\section{Traveling waves, solitary pulses and chemical turbulence}

It is well known that when the local kinetics of a reaction-diffusion system
has a limit cycle in phase space, the extended
system can support traveling wave trains \cite{26}.
In the special case of local kinetics of the form (2.1) ($\lambda $-$\omega $ systems),
Kopell and Howard \cite{26} have shown
that, for small wave numbers and small amplitude wave trains, the spatial shape of wave trains follows the temporal
behavior of the local coordinates
along the limit cycle in  phase space. This phenomena is
usually associated with the self-oscillatory characteristics of the local kinetics, and
therefore to target patterns and pacemakers \cite{4}.

The simplest way to obtain numerically stable traveling wave solutions in reaction-diffusion systems
with self-oscillatory dynamics, is to prepare
the extended system with initial conditions at the unstable steady state of the local system and  perturb it \cite{27}. In this case,  a traveling wave spreads from
the perturbation point. However, if the initial state of the extended system has values on the limit cycle, any sufficiently small perturbation of the extended system dies out and,
after some transient time,
the system oscillates globally.

We now investigate the persistence of wave type phenomena when the SNH bifurcation is
parametrically crossed.

Introducing diffusive coupling into model equation (\ref{E:9}), we obtain the system
of nonlinear parabolic partial differential equations,
\begin{eqnarray}\label{E:12}
\frac{\partial X}{\partial t} &=& D_X \nabla^2 X+\nu X - \beta Y \nonumber\\
&+& (X^2  + Y^2 )\left(a X - (b  + \frac{\alpha}{2}(1 - \frac{X}{\sqrt {X^2  + Y^2 } }))Y\right)\nonumber\\
\frac{\partial Y}{\partial t} &=& D_Y\nabla^2 Y+\beta X + \nu Y \nonumber\\
&+& (X^2  + Y^2 )\left(a Y + (b  + \frac{\alpha}{2}(1 - \frac{X}{\sqrt {X^2  + Y^2 } }))X\right) \nonumber\\
\end{eqnarray}
where $\nabla^2 =\left( \frac{\partial^2}{\partial x^2}\right)$, or
$\nabla^2 =\left( \frac{\partial^2}{\partial x^2}+\frac{\partial^2}{\partial y^2}\right)$ is the  Laplacian operator, and $D_X$ and $D_Y$ are diffusion coefficients. In the
following we always consider the case $D_X=D_Y>0$.

To analyze the development of wave type phenomena when we cross
the SNH bifurcation, we integrate numerically equation
(\ref{E:12}), using the bench-marked explicit integration method
developed in Dil\~ao and Sainhas \cite{27}, where optimal
convergence to the solution of the continuous system is achieved
if $\left( \Delta t  \max\{D_X,D_Y\}/(\Delta x)^2\right)=1/6$. We
consider that the reaction-diffusion equation (\ref{E:12}) is
defined over a finite one-dimensional or two-dimensional circular
domain with zero flux boundary conditions. Both domains have the
dimensional space length,
\begin{equation}\label{E:13}
L=N\sqrt{6\Delta t  \max\{D_X,D_Y\}}=N\Delta x
\end{equation}
where $N$ is the number of lattice sites along each spatial direction.
To fix parameters, we take equal diffusion coefficients, $D_X=D_Y=2\times 10^{-5}$, $a=-1$, $\nu=1$, and $b=1$. For these parameter values, by (\ref{E:8}), the SNH bifurcation
occurs for $\alpha_S=-\beta -1$.

To validate the numerical method, we have made several simulations for decreasing values of the integration time step $\Delta t$, comparing concentration profiles along an one-dimensional domain.
For integration times $\Delta t <0.02$ and during a finite time interval, the profile of the computed solutions coincide over
the one-dimensional domain, within a maximum global error in the concentrations of the order of $10^{-2}$.  Decreasing $\Delta t$, the global numerical error also decreases, insuring convergence. In all the simulations, we have used  the time step
$\Delta t=0.01$, and $\Delta x=1.095\times 10^{-3}$, calculated from  (\ref{E:13}).

Concerning initial conditions, we choose uniform concentrations
of $X$ and $Y$ along the one-dimensional and two-dimensional extended region with a
perturbation at the center of the domain. Then, at and beyond the SNH bifurcation ($\alpha <\alpha_S$),
we take two particular cases. In the first case, we consider that the initial state
of the two variables $X$ and $Y$ take the values of the unstable focus of the local kinetics
along the extended system, and we introduce a local perturbation at the center of the extended domain. The values of the $X$ and $Y$ perturbed values are
the coordinates of  the stable ($\alpha <\alpha_S$) or semistable ($\alpha =\alpha_S$) steady state  of the local kinetics.

In the
second case, the initial  state of the extended system take the values of  the stable or
semistable steady state of the local kinetics, and the local
perturbation in $X$ and $Y$ take the values of the unstable focus of the local kinetics.
For zero flux boundary conditions,
if we choose a steady state of the local kinetics as an initial state of the extended system, this steady state is also a solution
of the reaction-diffusion equations, and therefore, for a finite time interval, the presence of boundaries have no effect on the time evolution of the concentrations near the region of the
local perturbation. Under these conditions, all the subsequent analysis will be valid for finite and infinite domains.

The numerical integration of equation (\ref{E:12}) shows that if the local system in (\ref{E:12}) is
self-oscillatory  ($\alpha >\alpha_S$), and if initial conditions correspond to uniform
concentrations taken at the unstable steady state, with a perturbation away from it,
the system develops a traveling wave train
propagating along space.
However,  due to zero flux boundary condition, after a long integration time,
the whole extended system develops bulk oscillations.
A similar situation has been reported by Auchmuty and Nicolis \cite{28} for the Brusselator
model.

\begin{figure}[htbp]
\centerline{\includegraphics[width=3.96in,height=4.62in]{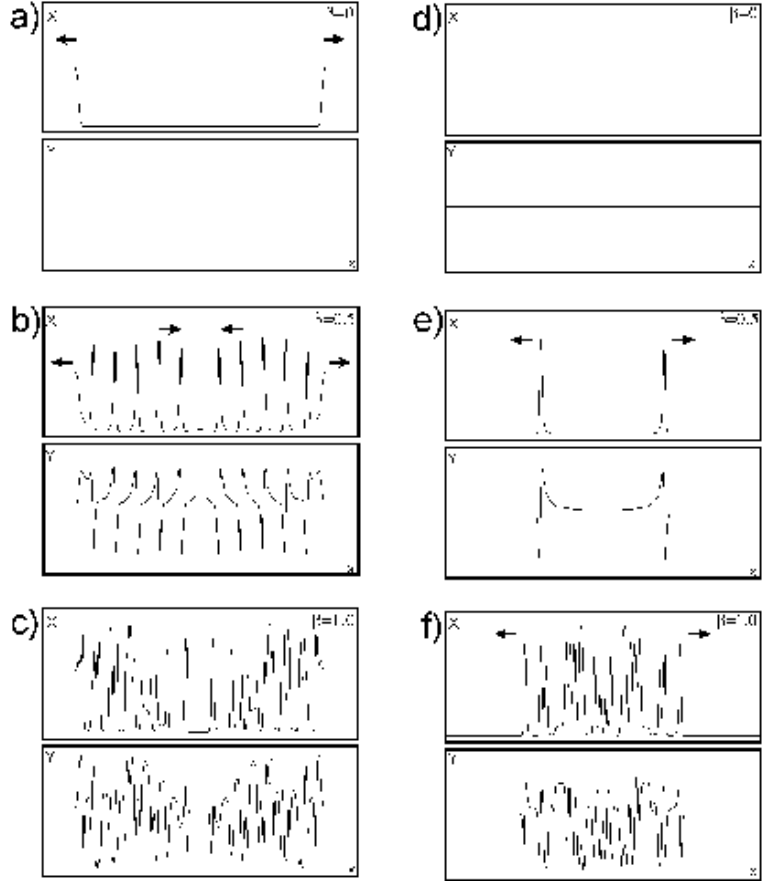}}
\caption{Solutions $X(x,t)$ and $Y(x,t)$ of the reaction-diffusion equation (\ref{E:12}) at the SNH bifurcation, in a one-dimensional domain with zero flux boundary conditions.
Fixed parameter values are: $D_X=D_Y=2\times 10^{-5}$, $a=-1$, $\nu=1$,  and $b=1$. The $\alpha $ and $\beta $ parameter values are: a) and d), $\alpha =\alpha_S=-1$, $\beta=0$;  b) and e), $\alpha =\alpha_S=-1.5$, $\beta=0.5$; c) and f), $\alpha =\alpha_S=-2.0$, $\beta=1.0$.
The total  integration time is $t=100$, for the time step  $\Delta t=0.01$.  The one-dimensional region has $N=2000$ lattice sites. In a), b) and c), the initial values of $X$ and $Y$ are taken at the unstable focus over the one-dimensional lattice except at the center of the domain, where $X$ and $Y$ assume the values of the semistable fixed point in phase space. In d), e) and f), the initial values of $X$ and $Y$ are taken at the semistable fixed point over the lattice, except at the center of the domain, where the system has been perturbed to the unstable focus with
a finite width perturbation of 160 lattice sites.
Propagating fronts, traveling waves and aperiodic regimes exist at the SNH bifurcation of the local dynamics.}\label{f3}
\end{figure}

For parameter values corresponding to
the SNH bifurcation of the local dynamics ($\alpha=\alpha_S=-\beta-1$), we have a semistable steady state with a well defined threshold, Fig.~\ref{f1}a) (and \ref{f1}c)).
Following the same strategy as in self-oscillatory systems, we prepare the initial state of
the extended system at the unstable
fixed point at the origin of phase space, and perturb it locally to the semistable state.
In this case, the solutions of the reaction-diffusion system (\ref{E:12}) develop fronts, periodic traveling waves and
aperiodic signals, depending on the parameter $\beta$, Fig.~\ref{f3}a)-c).
For $\beta =0$, Fig.~\ref{f3}a), the threshold of excitability coincides with the negative part of $X$ axis, and the system simply develops a propagating front. By (\ref{E:11}), increasing $\beta $, the front becomes a moving source initiating traveling waves. These traveling waves propagate in the opposite direction of the moving front, Fig.~\ref{f3}b).
For even larger values of $\beta $,
the waves generated by the moving front become aperiodic,  Fig.~\ref{f3}c).

Changing the initial conditions for an initial state at the semistable fixed point and
a perturbation of large width (160 lattice sites) with values at the unstable focus at the origin
of phase space, the reaction-diffusion system evolves in time in a completely different
way, Fig.~\ref{f3}d)-f). For $\beta =0$, Fig.~\ref{f3}d), the reaction-diffusion system evolves to the semistable
state. For larger values of $\beta $, the reaction-diffusion system develops a propagating pulse, as observed in
experimental systems, Fig.~\ref{f3}e), and for even larger values $\beta $ an aperiodic state
develops, Fig.~\ref{f3}f).

The comparison between cases a)-c) and d)-f) of Fig.~\ref{f3} leads to the conclusion that
target patterns, as observed in two-dimensional reaction-diffusion systems,
are not necessarily related with the oscillatory characteristics of the local systems.

\begin{figure}[htbp]
\centerline{\includegraphics[width=3.63in,height=2.29in]{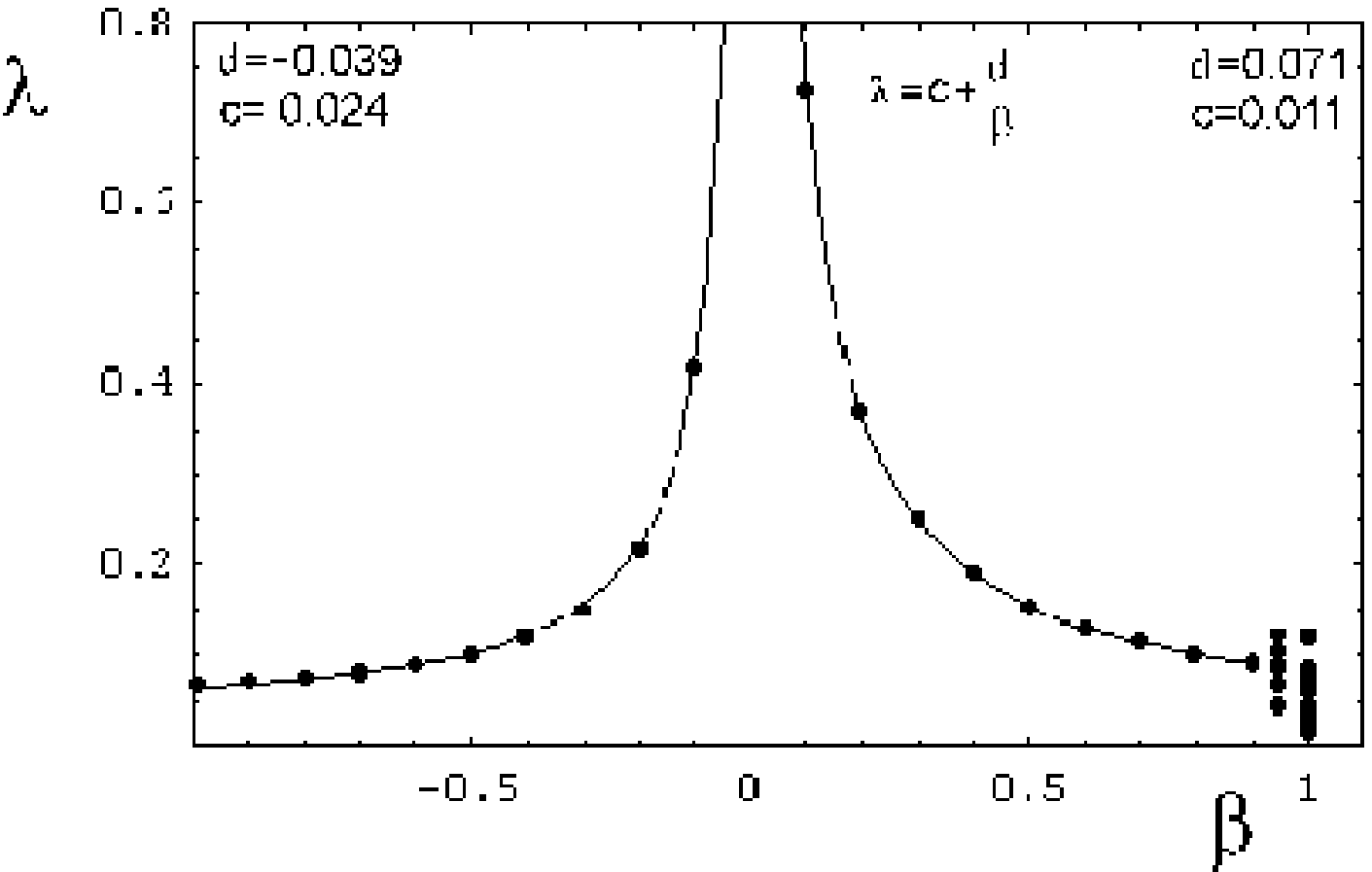}}
\caption{Wave length of traveling waves as a function of $\beta$,
taken at the SNH bifurcation ($\alpha=\alpha_S=\beta-b$), and the same
parameter values of Fig.~\ref{f3}.  For $\beta >0.95$, the front develops an aperiodic signal, usually identified as chemical turbulence. The variation of the wave length $\lambda $ with
$\beta $ follows the approximate law, $\lambda =c+d/\beta $. Numerically, $d=0.071$ and $c=0.011$, for $\beta >0$, and $d=-0.039$ and $c=0.024$, for $\beta <0$.}\label{f4}
\end{figure}

To analyze the
properties of periodic wave trains at the SNH bifurcation,
we have calculated numerically the wavelengths $\lambda $ between consecutive maxima along the extended media, Fig.~\ref{f4}.
For increasing values of $\beta $,  the aperiodic regime is reached. The wavelength as a function
of $\beta $ behaves as $\lambda \simeq a+d/\beta $.
At the SNH bifurcation, the stability of
wave trains seems to follow the same criteria as in self-oscillatory systems, where,
for increasing wave number $k=1/\lambda$,
the stability of periodic wave trains is lost \cite{4,26}, being followed by an aperiodic regime.
At the aperiodic regime, the distribution of wavelengths becomes continuous. In the literature of chemical
kinetics, this phenomena is  called chemical phase turbulence \cite{24}.

\begin{figure}[htbp]
\centerline{\includegraphics[width=3.98in,height=4.63in]{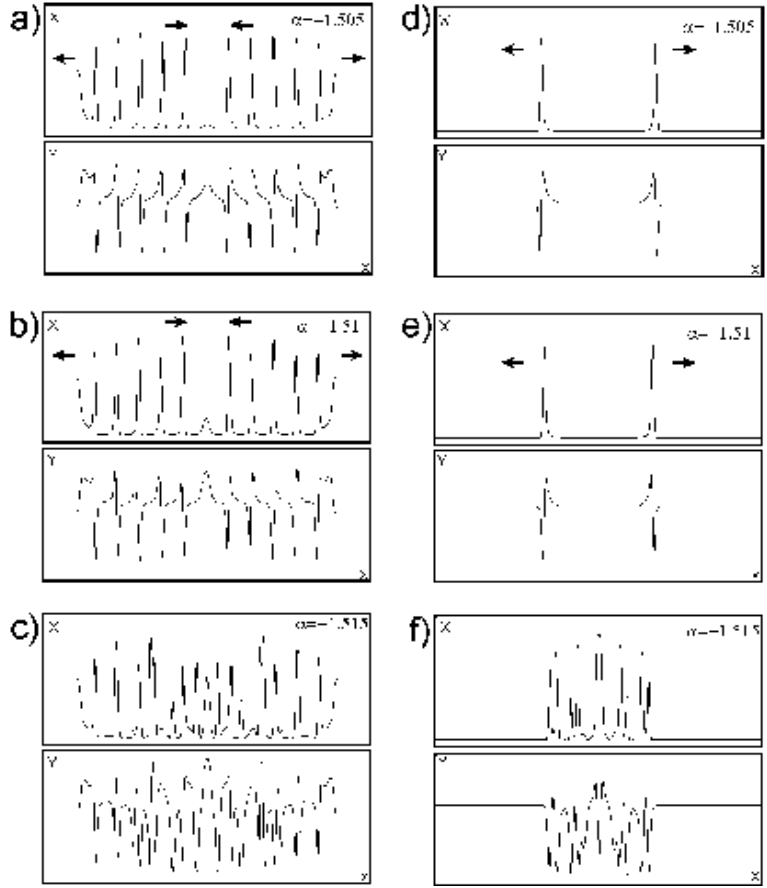}}
\caption{Solutions $X(x,t)$ and $Y(x,t)$ of the reaction-diffusion equation (\ref{E:12}) after SNH bifurcation,  in the same conditions of Fig.~\ref{f3}.
In a), b) and c), the initial values of $X$ and $Y$ are taken at the unstable focus over all the one-dimensional lattice except at the center of the domain, where $X$ and $Y$ assume the values of the stable fixed point in phase space. In d), e) and f), the initial values of $X$ and $Y$ are taken at the stable fixed point over the lattice except at the center of the domain, where the system has been perturbed to the unstable focus with
a finite width perturbation of $160$ lattice sites.
Fixed parameter values are: $D_X=D_Y=2\times 10^{-5}$, $a=-1$, $\nu=1$, $b=1.0$, $\beta =0.5$. The $\alpha $ parameter values are: a) and d), $\alpha =-1.505$; b) and e), $\alpha =-1.51$; c) and f) $\alpha =-1.515$. For $\alpha \approx -1.515 $, the generated wave front becomes aperiodic.
For $\alpha <\alpha_S $,  waves only persist in a narrow $\alpha$ parameter range.}\label{f5}
\end{figure}

To analyze the persistence of wave type solutions when the SNH bifurcation is crossed, we have fixed $\beta =0.5$.
In this case, the SNH bifurcation occurs for $\alpha_S=-\beta-1=-1.5$, and we have investigated the limits in $\alpha $ where periodic traveling waves persist, Fig.~\ref{f5}. The $\alpha $ parameter controls the distance from the stable steady state to the threshold of excitability.
After the SNH bifurcation, in a range of the $\alpha$ parameter $\alpha \in (-1.515,-1.5]$, traveling wave solutions persist, Fig.~\ref{f5}a).
The initial perturbation develops a moving source initializing traveling waves, which propagates in the opposite direction of the front.
When $\alpha $ approaches the value $\alpha= -1.515$, the amplitude of different wave spikes changes, but the
wavelength remains constant, Fig.~\ref{f5}b). Around $\alpha \approx -1.515$, the propagating signal becomes aperiodic, Fig.~\ref{f5}c). Decreasing even further $\alpha $, the turbulent regime behind the moving source decreases. At low $\alpha $ values, the moving source
is followed by a narrow zone where the generated pulse disappears.

Therefore, after the SNH bifurcation,   wave propagating phenomena exists in a narrow $\alpha $ parameter range. Geometrically, this parameter range depends of the proximity
of the saddle and node pair of fixed points.

For an initial state in the unstable focus and a perturbation at the stable fixed point in the phase space, Fig.~\ref{f5}d)-f), depending of the width of the perturbation, the perturbation dies out, or the extended system develops a finite train of propagating pulses, or develops an aperiodic regime. In the case of the simulations of Fig.~\ref{f5}d), we obtain a traveling pulse for perturbation radius larger than $4$ lattice sites. For smaller perturbation sizes, the perturbation dies out.
For large perturbation widths and decreasing $\alpha $, the isolated pulses develop and
propagate
in the same region  $\alpha \in (-1.515,-1.5]$, Fig.~\ref{f5}d)-e).

\begin{figure}[htbp]
\centerline{\includegraphics[width=2.79in,height=3.63in]{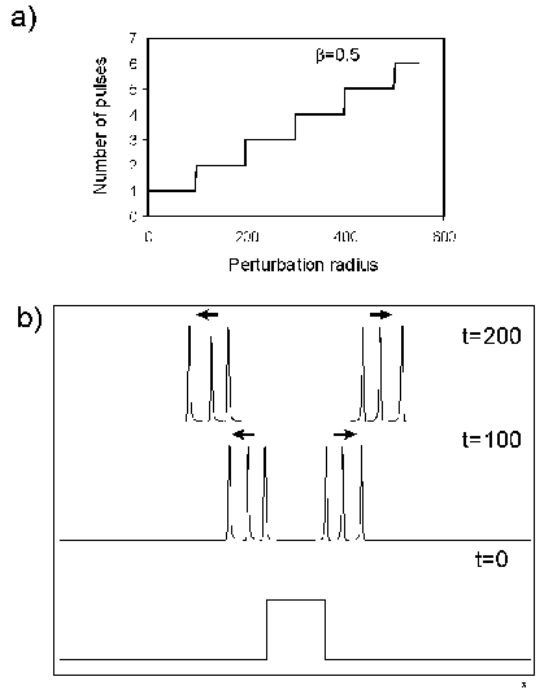}}
\caption{a) Number of traveling pulses as a function of perturbation radius at the SNH bifurcation. The initial values of $X$ and $Y$ are taken at the semistable fixed point over all the one-dimensional lattice except at the center of the domain, where the system has been perturbed to the unstable focus with
a variable width perturbation.
b) Time evolution of an initial conditions leading to two packets of three propagating pulses. In this case the perturbation radius has 250 lattice sites.
All the simulations were carried out in an one-dimension domain with $N=4000$ lattice sites. The parameters are the same as of Figs.~\ref{f3}b) and ~\ref{f3}e), with $\beta=0.5$ and $\alpha =\alpha_S=-1.5$.}\label{f6}
\end{figure}

 We have analyzed the dependence on the initial perturbation size. Increasing the radius of the perturbation, more and more traveling pulses appear, Fig.~\ref{f6}a). The perturbation increments, excluding the first one, are equal. For the parameter values of Fig.~\ref{f6}, the
media can propagate waves with a characteristic wavelength of 141 lattice sites (Fig.~\ref{f4}), implying that for an increment in the perturbation width of 110$\%$ of the wave length, a new propagating pulse appears, and we have a moving pulse packet. In Fig.~\ref{f6}b),
we show the time evolution of an initial condition leading to two packets of three
propagating pulses.
For this particular choice of initial conditions, if the perturbation is not
exactly in the unstable focus, then the system can only develop one pulse or the perturbation dies out.
This situation occurs for perturbation values all over the phase space, even using large perturbation radius.

The phenomena of persistence of traveling wave trains and pulses at and after the SNH bifurcation has a simple explanation.
At the SNH bifurcation ($\alpha=\alpha_S$), when the extended system has initial conditions in the unstable focus, any perturbation from it approaches the semistable
fixed point. Driven by diffusion, the points spatially close to the perturbation generate a line in phase space. For $\beta \not=0$, this line always intersects the threshold of excitability
in an infinite number of points and a traveling wave is generated. When the extended
system has initial conditions in the semistable fixed point,
any small fluctuation on the concentrations of $X$ and $Y$ crossing the threshold
drives nearby points of the extended system
to trajectories in phase space with different lengths. In this case, we have pulse generation.
After the SNH bifurcation, if the distance between the stable steady state and the saddle point is small, larger perturbations cross the excitability threshold,
and an
instability  in space develops. As $\alpha $ decreases
away from $\alpha_S$, the distance in phase space between the stable steady state and the saddle point increases, and the diffusion driven instability disappears.

\begin{figure}[htbp]
\centerline{\includegraphics[width=3.91in,height=2.10in]{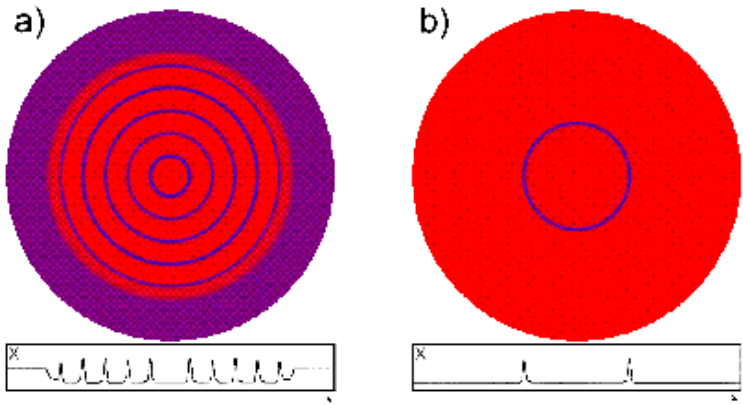}}
\caption{Solutions $X(x,t)$  of the reaction-diffusion equation (\ref{E:12})  in a two-dimensional  domain with zero flux boundary conditions.  The local dynamics has a saddle node pair of fixed points and
an unstable fixed point at the origin of coordinates.
Blue represents higher concentrations of $X$ and red lower concentrations. We also show the spatial variation of the concentration of $X$
along  a cross section passing through the center of the two-dimensional domain.
In a), the initial values of $X$ and $Y$ are taken at the unstable focus over all the one-dimensional lattice except at the center of the domain, where $X$ and $Y$ assume the values of the stable fixed point in phase space. In b), the initial values of $X$ and $Y$ are taken at the stable fixed point over the lattice except at the center of the domain, where the system has been perturbed to the unstable focus with
a finite width perturbation of 160 lattice sites. In
a), we show the formation of a traveling wave.
In b), we show a propagating solitary pulse.  The two-dimensional circular domain is embedded in a two-dimensional square lattice with $2000$ by $2000$ sites.
The parameter values are: $D_X=D_Y=2\times 10^{-5}$, $a=-1$, $\nu=1$, $b=1.0$, $\beta =0.5$ and $\alpha =-1.51<\alpha_S$, as in the cases of Figs.~\ref{f5}b) and ~\ref{f5}e). In both figures, the  total integration time  is $t=100$.
}\label{f7}
\end{figure}

In two dimensional domains the situation is similar to the one dimensional case. For
example, preparing the system as in the cases of Figs.~\ref{f5}b) and ~\ref{f5}e),  we obtain a
traveling wave train and a traveling isolated pulse, respectively, independently of the
parameters of the local dynamics, Fig.~\ref{f7}a)-b).
Further numerical simulations for the two-dimensional case shows that the excitable
local system coupled with diffusion in the two phase space variables
can develop spirals.
However, in the two-dimensional case, after the SNH
bifurcation, the $\alpha $ parameter range where propagating wave trains and pulses persist differs
from the one-dimensional case.

\section{Conclusions}

We have obtained a local model of an excitable system exhibiting a Hopf and a saddle-node homoclinic (SNH) bifurcation. This model has been constructed introducing a phase dependence on the angular coordinate of the versal unfolding of the Hopf bifurcation. The bifurcation parameters of the model are completely independent.
In extended media, the excitability state occurs when the SNH bifurcation of the local model is crossed.
The excitable local system has a heteroclinic orbit connecting two unstable fixed points, and this orbit is close to a stable steady state.

We have shown that in excitable extended systems with a sharp threshold and coupled by diffusion,  traveling waves, fronts, isolated pulses, packets of pulses and aperiodic wave trains develop.
In all the cases, we have taken equal diffusion coefficients for the local variables of the model.
Depending on the initial conditions of the extended system,
traveling periodic waves, isolated pulses and packets of pulses exist for the same parameter values of
the excitable system.
Propagating pulses and packets of pulses only exist at and beyond the SNH bifurcation of the local model. Traveling wave trains exist on both sides of the SNH bifurcation, but the wave lengths change discontinuously when the SNH bifurcation is crossed.

Excitable systems  with a precise threshold and equal diffusion coefficients can show  a large variety of behaviors, as found in real chemical systems: Isolated propagating pulses and fronts, as in the iodate-arsenous acid extended system \cite{13}, packets of propagating pulses or bursting, as in neurons and pancreatic $\beta $-cells \cite{29}, and chemical turbulence \cite{24}. Excitable systems with precise threshold can also show target patterns similar to those observed in the Belousov-Zhabotinskii experiment \cite{30}. In the case of  the Belousov-Zhabotinskii reaction,  if we assume that the local kinetics is self-oscillatory and can be described by a two variable model \cite{23}, only large perturbations for the vicinity of an unstable steady state can be in the origin of  the emergence of target patterns. These perturbations can be seen as the local
fluctuations of  concentration occurring in the media at the microscopic level. This could explain way pacemakers appear sparsely and spontaneously in experiments in the extended media.

The relation between chemical turbulence associated to the appearance of an aperiodic regime as reported here,  are dependent of the parameters of the local kinetics and are less dependent of the initial distribution of the initial states in phase space.

The model presented here
shows qualitatively the known types of topologically distinct activity waves as observed in real chemical experiments, and the different activity waves can be
tuned by different initial conditions, maintaining the parameters of the local kinetics fixed.

Finally, one of the consequences of the approach presented here is that in excitable extended systems with sharp threshold (Type II excitability), we can have pulse packets with any number of  bursts, whereas in models with finite width threshold (Type I excitability), packets of propagating pulses have never been observed.
This enables to design experiments in order to decide between the two types of excitability.

\section*{Acknowledgments}
We would like to thank Zolt\'an Noszticzius for stimulating discussions.
This work has been partially supported by the POCTI Project P/FIS/ 13161/1998 (Portugal). One of us (A. V.) acknowledges  a traveling grant from the REACTOR program of European Science Foundation.

\label{end}
\medskip

 \medskip

\end{document}